# Semimetal-Monolayer Transition Metal Dichalcogenides Photodetectors for Wafer-Scale Ultraviolet Photonics

Hon-Loen Sinn,[1] Aravindh Kumar,[2] Eric Pop,[2,3,4] and Akm Newaz[1]

[1]Department of Physics and Astronomy, San Francisco State University, San Francisco, California 94132, USA

[2]Department of Electrical Engineering, Stanford University, Stanford, California 94305, USA

[3]Department of Materials Science and Engineering, Stanford University, Stanford, California 94305, USA

[4]Precourt Institute for Energy, Stanford University, Stanford, California 94305, USA

Emails: hsinn@sfsu.edu, akumar47@stanford.edu, epop@stanford.edu, akmnewaz@sfsu.edu

## Abstract

Atomically thin two-dimensional (2D) transition metal dichalcogenides (TMDs), such as $MoS_2$, are promising candidates for nanoscale photonics because of strong-light matter interactions. However, Fermi level pinning due to metal-induced gap (MIGS) states at the metals-monolayer $MoS_2$ interface limits the application of optoelectronic devices based on conventional metals because of the high contact resistance of the Schottky contacts. On the other hand, a semimetal-TMD-semimetal device can overcome this limitation, where the MIGS are sufficiently suppressed and can result in ohmic contacts. Here we demonstrate the optoelectronic performance of a bismuth-monolayer (1L) $MoS_2$-bismuth device with ohmic electrical contacts and extraordinary optoelectronic properties. To address the wafer-scale production, we grew full coverage 1L $MoS_2$ by using chemical vapor deposition method. We measured high photoresponsivity of 300 A/W in the UV regime at 77 K, which translates into an external quantum efficiency (EQE) ~ 1000 or $10^5$%. We found that the 90% rise time of our devices at 77 K is 0.1 ms, which suggests that the current devices can operate at the speed of ~ 10 kHz. The combination of large-array device fabrication, high sensitivity, and high-speed response offers great potential for applications in photonics that includes integrated optoelectronic circuits.

## 1. Introduction

Ultraviolet photodetectors (UVPDs) are at the heart of many applications ranging from biological analysis, environmental sensors, fire monitoring, and space exploration to ultraviolet (UV) radiation detection[1-5]. Photomultiplier tube UVPD is commonly used because of their speed and sensitivity (single or few photons in the UV). However, photomultipliers are bulky, large in size (2-3 cm), and require high bias voltages (~ 700 V). These characteristics are not suitable for remote and arrays of sensors (e.g., environmental monitoring, space exploration, UV astronomy, UV imaging camera) because of weight, size, and power constraints. For these types of applications, small pixel sizes and low power operation are requirements that need nanoscale solutions.

Many materials have been explored to develop efficient ultraviolet PDs. Two extensively researched materials are group III-nitride semiconducting materials,[2, 4] including AIN, AlGaN, InAlGaN, and group IIB semiconductors[1], such as ZnO. Group III-nitride based semiconductors have several drawbacks for their use as reliable UVPDs: they have significant lattice mismatch and thermal coefficient mismatch with Si[6], which lead to poor crystal quality and the formation of cracked networks[7-9]. Group IIB ZnO semiconducting materials are attractive choices as they have a wide band gap (~3.7 eV), a high exciton binding energy (~60 meV), and high chemical stability.[1, 10-15] However, the high surface-to-volume ratio of ZnO nanostructures, their grain boundaries, low mobility, and very long response time[16-24] severely impact detector performance. Silicon-based UV PDs have major limitations: their band gap is 1.1 eV with a maximum responsivity around 700 nm, which drops dramatically in the UV range (responsivity ~0.3 AW$^{-1}$ at 400 nm[25]).

Two-dimensional transition metal dichalcogenides (TMDs), such as $MoS_2$, materials provide an attractive platform for nanoscale photonics due to their atomic scale thickness, strong-light matter interaction, and favorable mechanical, and electrical properties.[26-31] Photons impinging on a monolayer TMDs (1L-TMDs) will produce a direct band-gap optical transition, also known as the *A* and *B* transitions. [26-31] In addition, there also exists a pair of van Hove singularity (vHS) assisted excitonic transitions (referred to as *C*[32] and *D* peaks[33]) in the UV regime above ~(3 - 4 eV) eV.[32] These vHS singularity excitons cause extraordinarily high photon absorption of 1L-TMDs (~40% for 1L-MoS$_2$[34-36]). These *vHS* excitons form within the continuum of the quasiparticle state, i.e., above the band edge, and decay spontaneously. [32] Due to spontaneous decay, the lifetime of the *C*-/D-excitons is short ($\tau_C \sim 0.4$ ns).[37] High absorption in the UV and shorter lifetime of the vHS excitons create an opportunity to develop an efficient nanoscale UV photodetector.

However, the formation of metal-induced gap states (MIGS) at the metal-semiconductor interface causes an energy barrier—which leads to high contact resistance, non-linear current-voltage (*I-V*) characteristics, and poor current delivery capability.[38-40] All these factors limit the use of TMDs as next-generation photonic devices. Recently, it has been reported that a semimetal-TMDs-semimetal (STMDS) device can overcome this limitation, where the MIGS are sufficiently suppressed, which results in creating ohmic contacts.[41, 42] The formation of ohmic contacts may improve extraordinarily optoelectrical performances of an STMDS device and may find a wide range of applications in the next-generation device applications. However, the optoelectronic properties of an STMDS device have not been demonstrated yet.

In this work, we demonstrate the optoelectronic properties of an STMDS device based on bismuth (Bi)-monolayer MoS$_2$-Bi photodetector devices. For wafer-scale applications of STMDS, full-coverage growth of monolayer TMDs (1L TMDs) is critical. To address this, we grew monolayer MoS$_2$ (1L MoS$_2$) by solid-source chemical vapor deposition (CVD). We measured a high photoresponsivity of 300 A/W in the UV regime at 77 K, which translates into an external quantum efficiency (EQE) ~ 1000 or 10$^5$%. By measuring the photocurrent spectroscopy, we found that our devices are most sensitive in the UV range. We have found that the 90% rise time and fall time of our devices is 0.1 ms, which suggests that the current devices can operate at the speed of ~ 10 kHz.

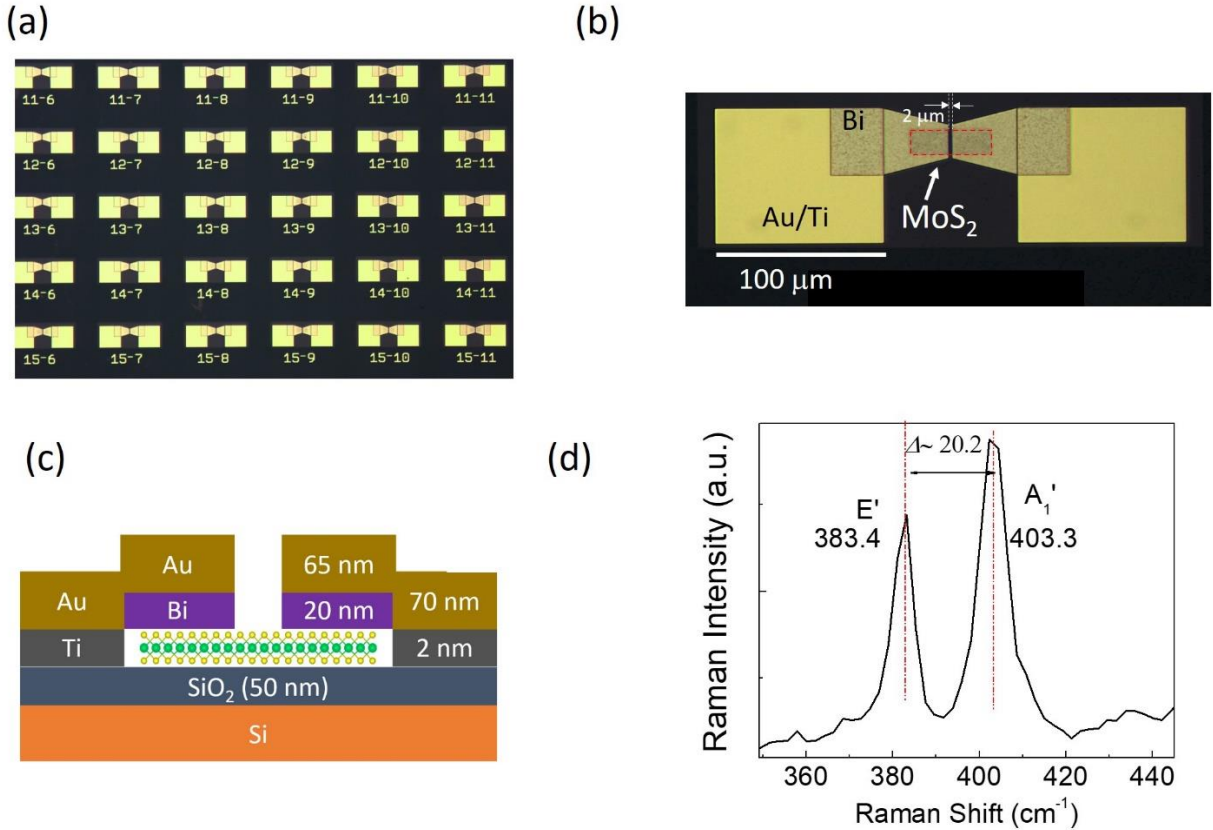

**Figure 1:** Device Characteristics. (a) Optical image of arrays of samples. (b) Blow up image of a single device. First, a full-area coverage monolayer MoS₂ was grown by solid source chemical vapor deposition (CVD). The sample strips were prepared by optical lithography followed by oxygen plasma etching (red-dashed line rectangle). (c) The device structure is shown schematically. (d) Raman spectroscopy from the monolayer MoS₂ sample. The presence of two peaks (E' and $A_1'$) confirming the monolayer nature of CVD grown MoS₂. The excitation laser source was 532 nm.

## 2. Results and Discussion

We fabricated the TMD device on SiO₂/undoped-Si wafer. We selected an undoped Si wafer, obtained from a commercial vendor (University Wafer Inc.) to reduce the photogating effect, which arises due to the accumulation of the photogenerated carrier at the interface between SiO₂ and Si that gates the TMD electrostatically.[32, 43, 44] The surface crystal orientation of the undoped wafer was <100> and the resistivity of the wafer was >20,000 ohm-cm. A 50 nm thick $SiO_2$ was grown thermally, followed by growing monolayer $MoS_2$ by solid-source CVD technique. A detailed description of the TMD layer preparation is available in the work by Smithe *et al.*[45] The optical image of an array of devices is shown in Fig.1a. The blown-up view of one device is shown in Fig.1b. A rectangular strip of 1L $MoS_2$ of size 20 $\mu$m$\times$ 50 $\mu$m was patterned using optical lithography followed by $O_2$ plasma etching. The red-dashed line rectangle in Fig.1b shows the strip of 1L $MoS_2$. Next, a semimetal (bismuth (Bi)) contact was formed by using optical lithography followed by electron beam (e-beam) evaporation. Finally, large square wire-bonding contact pads (size: 100 $\mu$m$\times$ 100 $\mu$m) were prepared using optical lithography followed by thermal evaporation of Ti (2 nm) and Au (70 nm). The dimension of different components of a device is shown in Fig.1c. The monolayer nature of the film was confirmed by Raman spectroscopy measured at room temperature using a homemade system. Confocal micro-Raman measurements were performed after completing the device fabrication. A 100$\times$ objective lens with a numerical aperture of 0.85 was used. The excitation source was a 532 nm laser (2.33 eV) with an optical power of $\sim$500 $\mu$W. The Raman spectrum of the sample is shown in Fig.1d, which shows two signature peaks ($E' = 383.4$ cm$^{-1}$ and $A'_1 = 403.3$ cm$^{-1}$) of 1L $MoS_2$. The gap between the Raman peak is $\Delta = 20.2$ cm$^{-1}$, which confirms that the sample is 1L $MoS_2$.[46]

To study the temperature-dependent electrical and optoelectronic properties of an STMDS sample, we mounted the samples inside a microscopy cryostat (Janis Research, ST-500) equipped with electrical feedthrough for electro-optical measurements. The cryostat was coupled with an Olympus microscope equipped with a long-working distance objective (magnification 40x). For wavelength-resolved measurements, we used a broadband light source (tungsten–halogen lamp) coupled to a double-grating monochromator (Acton Spectra Pro SP-2150i). The photocurrent was measured by employing the lock-in techniques.[47] The optical beam was modulated by an optical chopper ($f = 79$ Hz). The optical power on the sample was determined using a well-calibrated Si *p-i-n* photodetector (Hamamatsu S1223).

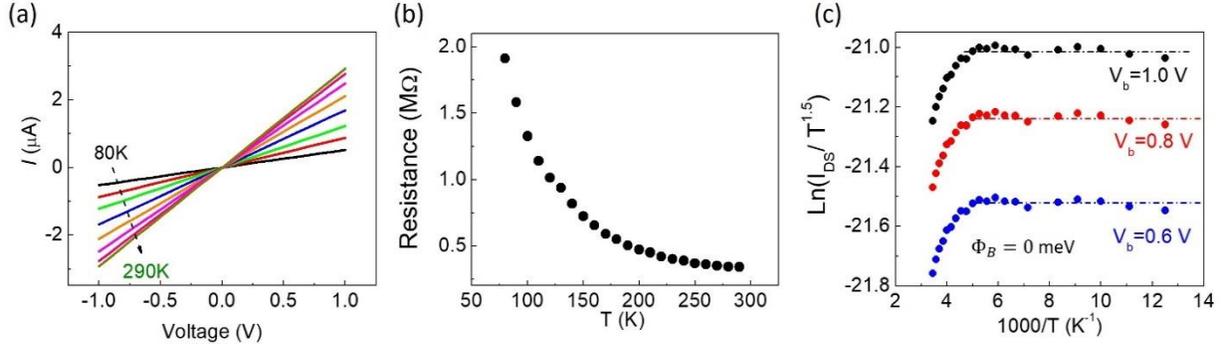

Figure 2: Electrical transport properties of a device. (a) Current-voltage (I-V) curve of a device at different temperature from 80 K to 290 K at every 30 K steps. The I-V curves were measured in the dark. The IV curve shows excellent ohmic behavior. (b) The resistance as a function of temperature. The resistance was measured from the slope of the I-V curves shown in Fig.(a). (c) Determination of the Schottky barrier height for different bias voltage ($V_b$). The Schottky barrier height vanishes at temperatures below 180 K. See the main text for details.

We measured the electrical transport properties of the sample at different temperatures from 80 K to 290 K as shown in Fig.2. We measured the current using a programmable source meter (Keithley 2400) connected to a current preamplifier (Stanford Research SR570). The voltage signal from the current amplifier was measured by using a digital multimeter (Keithley 2000).

We studied a total of six devices, all of which showed similar results. The current-voltage (*I-V*) at different temperatures from 80 K to 290 K is shown in Fig.2(a). We observed that the I-V curves demonstrate very linear behavior near the zero voltage, which suggests that the MIGS states are suppressed and the contacts are Ohmic in nature. The resistance at different temperatures from 80 K to 290 K at every 30 K step is shown in Fig.2(b). The temperature dependence of the resistance demonstrates very semiconductor-like behavior. We note that the contact resistance also changes as we increase the temperature.

The Schottky barrier can be determined from the temperature-dependent *I-V* characteristics. The current ($I$) through an atomically thin 2D 1L MoS$_2$ is governed by the 2D thermionic emission equation,[48] which employs a reduced $T^{1.5}$ power law for two-dimensional transport:[49]

$$I_{ds} = A_{2D}^* S T^{1.5} \exp\left[-\frac{q}{kT}\left(\Phi_B - \frac{V_b}{n}\right)\right] \tag{1}$$

where $A_{2D}^*$ is the 2D equivalent Richardson constant, $S$ is the contact area, $n$ is the ideality factor, $\Phi_B$ is the Schottky barrier, $k$ is the Boltzmann constant, $q$ is the electron charge, and $V_b$ is the voltage applied between the terminals. To determine the Schottky barrier, one can utilize the Arrhenius plot, i.e., $\ln\left(\frac{I}{T^{1.5}}\right)$ vs $1000/T$ as shown in Fig.2c.[49, 50] The slope of the plot will give $m = -\frac{q}{1000k}\left(\Phi_B - \frac{V_b}{n}\right)$. If we plot the slope as a function of $V_b$, the intercept of the new plot, $c_0 = -\frac{q\Phi_B}{1000k}$, will give a direct measure of the Schottky barrier height. We observed that the Arrhenius plots are horizontal below 180 K for different $V_b$ as shown in Fig.2c, which means that the slopes are zero or $m = 0$. Now, if we plot $m$ vs $V_b$ for temperatures below 180 K, the intercept of the plot $c_0$, will be zero, which means $\Phi_B = 0$. Hence, the Schottky barrier $\Phi_B$ of our devices vanishes below 180 K.

The optical and optoelectronic properties of a sample are shown in Fig.3. Photoluminescence spectrum was taken from the sample at room temperature by exciting using a 532 nm green laser and is shown in Fig. 3a. We observed two peaks at 675 nm and 620 nm, which correspond to the A- and B- excitons in 1L MoS$_2$,[32, 51, 52] confirming that our samples are of monolayer nature.

We measured photocurrent using two different optical sources; (i) a laser of wavelength 405 nm and (ii) a broadband tungsten-halogen thermal source. In atomically thin 2D TMD-based photodetectors, the photocurrent originates from two main mechanisms: (i) the photoconductive effect, where the photogenerated electron-hole pairs increase the carrier density and the electrical conductivity; and (ii) the photogating effect, where the photogenerated carriers filled the localized trap states and cause a shift of the Fermi energy.[53-59] We have probed the photoconductive mechanisms in our devices by measuring photocurrent with varying light power, bias voltages, and laser pulse width. We have found that photogating is the dominant mechanism in our devices.

To understand the photocurrent mechanism, we determined the wavelength-resolved photoresponsivity (photocurrent per unit power of lights), $R_\lambda$, of the sample for a wide range of wavelengths from 350 nm to 1050 nm. Fig.3b shows the photoresponsivity of an STMDS sample measured at 77 K. The light power on the sample was calibrated by a Si *p-i-n* photodetector (Hamamatsu S1223). We measured scanning photocurrent image of the sample to map the region of photocurrent contribution. We found zero photocurrent outside our sample (see Supporting Information).

We observed three important features. First, we observed one peak at 380 nm. We attributed this peak to the van Hove singularity exciton, which also causes the highest absorption of photons.[32] Second, there is a peak in the infrared region, whose origin is currently unknown. Since it has been observed that the defects in CVD-grown $MoS_2$ crystal can cause a wide peak in the infrared region,[60] we attribute this infrared peak to a defect-induced peak in the CVD-grown crystal. Third, the photocurrent peaks due to A- and B-excitons are not visible in the spectrum. We attribute this due to the photogating effect which is the main mechanism of the photocurrent generation in our STMDS devices. There are two competing photocurrents in our devices. One photocurrent is originating from the exciton dissociations[32] and the second photocurrent is originating from the photogating.[61] Since the photocurrent due to photogating is several order magnitudes higher than the photocurrent due to exciton dissociation, the signatures of A- and B- exciton peaks do not appear in the photocurrent spectrum. We observed peaks in the photocurrent spectrum due to the A- and B- excitons for samples that were grown following the similar solid source CVD method and were prepared on a quartz crystal substrate (see Supporting Information). Since the photogating effect is absent in a sample prepared on a quartz substrate, exciton dissociation is the dominant photocurrent mechanism and the exciton related peaks appear in the photocurrent spectrum. We presented details explanations of the photogating mechanism with supporting information below.

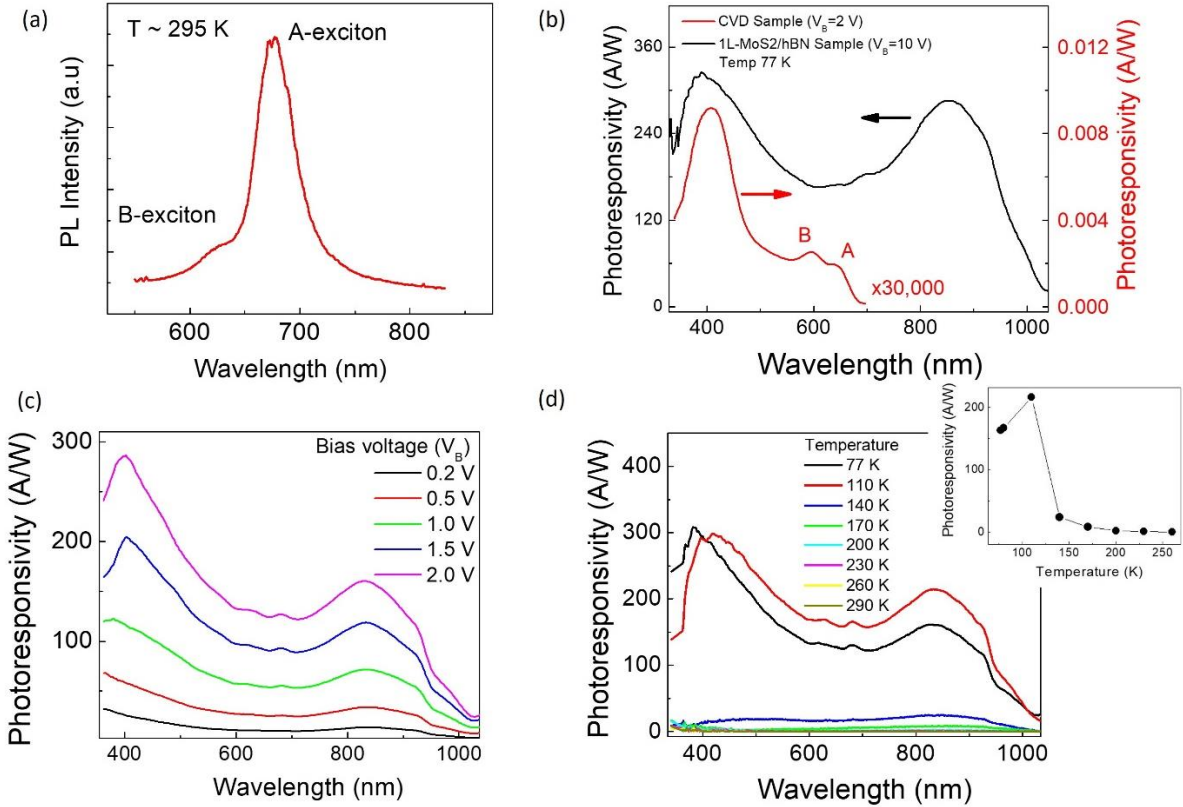

Figure 3: Optical and optoelectrical properties of a Bi-MoS$_2$-Bi device. (a) Photoluminescence spectrum from a 1L MoS$_2$ sample measured at room temperature. The excitation laser wavelength was 532 nm. Two neutral exciton peaks, A- and B-peaks, appear at 675 nm and 620 nm, respectively. (b) Photoresponsivity (current per unit light power) of the sample (black solid line) at different wavelength. The red solid line is the photocurrent spectroscopy from a monolayer Au-MoS$_2$-Au sample on a glass substrate encapsulated by hBN. The photoresponsivity in the Au-1L MoS$_2$-Au is 30,000 smaller than that from our Bi-MoS$_2$-Bi sample. Two peaks at 650 nm and 590 nm in the photoresponsivity spectrum of Au-1L MoS$_2$-Au sample are due to the A- and B-excitons and are marked by A and B, respectively. (c) Photoresponsivity of a sample at different bias voltages measured at 77 K. We found a linear behavior as we increased the bias voltage. (d) Photoresponsivity of a sample at different temperature. The photoresponsivity vanishes as the temperature is increased. Inset: the amplitude of the photocurrent peak in the infrared region (wavelength ~ 880 nm) at different temperatures.

To compare the photocurrent spectrum with a conventional micro-exfoliated TMDs, we also studied the photocurrent spectroscopy of a high-quality 1L MoS$_2$ sample encapsulated by a thin hexagonal boron nitride (hBN) flake as shown in the Fig.3b (red line). The samples were fabricated on a glass substrate using the dry transfer technique and were electrically connected to a prep-patterned Au electrode. For detailed results on fabrication and photocurrent spectroscopy of exfoliated 1L MoS$_2$, we guide the reader to our previous publications by Benson et al.[62] The right axis in Fig.3b presents the photoresponsivity from the 1L MoS$_2$ sample. The photoresponsivity in our STMDS samples is ~30,000 larger than that from a 1L MoS$_2$ sample, which suggests extraordinarily high photocurrent in STMDS device. The two peaks at 650 nm and 590 nm in the photoresponsivity spectrum of 1L-MoS2/hBN sample in Fig.3b are due to A- and B- excitons.

To compare the photocurrent spectrum to other metallic contacts to CVD grown MoS$_2$, we also studied solid state CDV grown 1L-MoS$_2$ samples with Ag/Au contact instead of Bi contact. We have observed that the photoresponsivity in MoS$_2$/Au devices is six orders of magnitude lower than that in MoS$_2$/Bi devices (see Supporting Information).

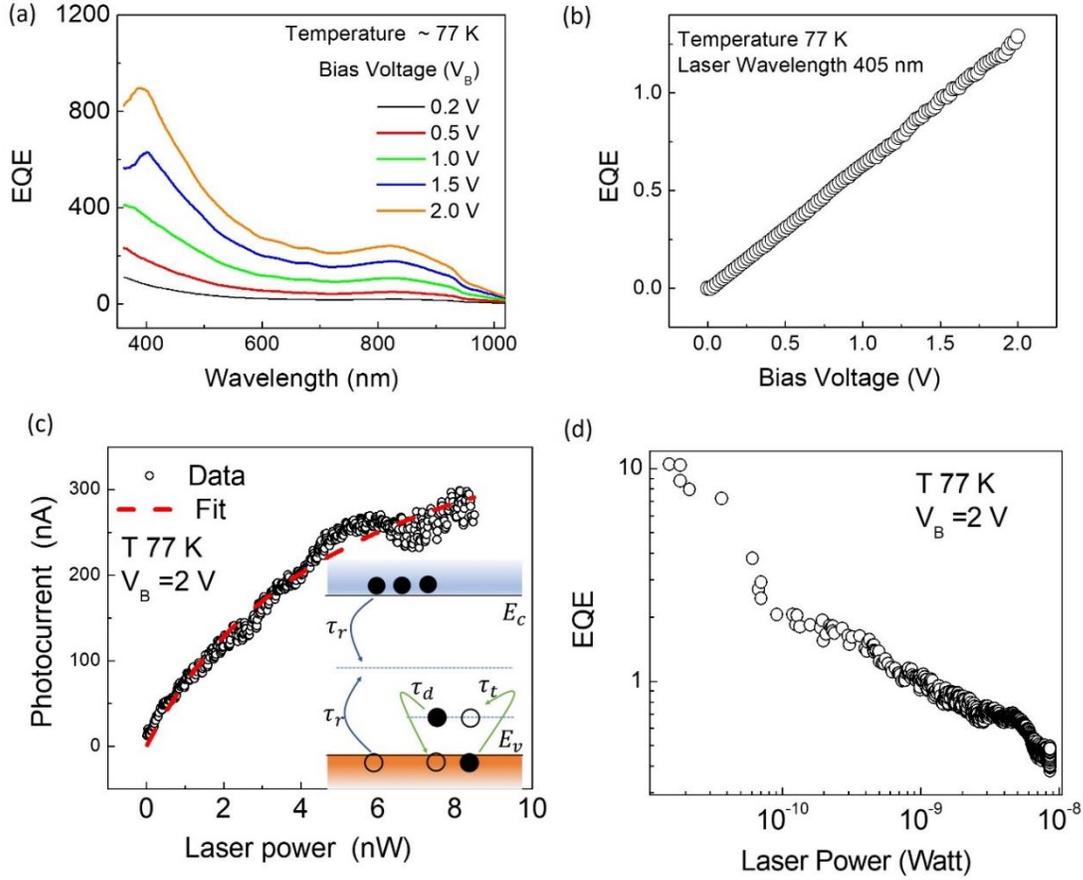

Figure 4: Measurements of external quantum efficiency (EQE), which is defined as the number of electrons for one photon. (a) EQE of a sample at different wavelength. We have observed very high EQE. The highest efficiency was observed for UV photons (~ 400 nm). Note that in the EQE calculation, we have not included the photon absorption efficiency. If we include the photon absorption efficiency, the EQE will be an order of magnitude higher. (b) The measurement of EQE as a function of bias voltage. Here we used a 405 nm laser. To improve the signal, we have used a high-power laser beam; that is why EQE is lower for this measurement. (c) Photocurrent as a function of a 405 nm laser power measured at 77 K. The bias voltage was 2 V. Inset: Simplified energy band diagram that shows the main features of the charge trapping and detrapping model. The valence band trail is approximated by a discrete distribution of hole traps with density $D_t$ (occupation of traps $p_t$). The holes are trapping into the states and de-trapping out of the sates with a rate $1/\tau_t$ and $1/\tau_d$, respectively. (d) EQE as a function of the laser power. We see that EQE decreases logarithmically as we increase the power, which is clear sign that the high EQE is originating from the photogating effect as describe above.

To understand the electric field-dependent photocurrent, we also studied photocurrent spectrums for different bias voltages as shown in Fig.3c. We see that photocurrent increases linearly as we increase the bias voltage. Fig.3d presents the temperature dependence property of the photocurrent spectrum at a wide range of temperatures from 80 K to 298 K. We observed that photocurrent decreases as we increase the temperature and almost disappears near room temperature ($T \geq 250$ K). To determine the temperature-dependent behavior of photocurrent, we measured the amplitude of the photocurrent peak in the infrared (~ 880 nm) at different temperatures as shown in the inset of Fig.3d.

To determine the photodetector performance, we determined the external quantum efficiency (EQE), which is defined as a ratio of the number of electrons in the external circuit to the number of incident photons. Fig.4a presents EQE for different samples, which is connected to photoresponsivity $R_\lambda$, by EQE $= \frac{R_\lambda}{\lambda} \times 1240$, where $R_\lambda$ is the responsivity in A/W, and $\lambda$ is the wavelength in nm. We observed an EQE of 1000 or $10^5$% at 77 K for a bias of 2V. Note that high photoresponsivity was reported for a 1L $MoS_2$ phototransistor sample before, where the samples are gated by a very large gate voltage.[61] To the best of our knowledge, we observed the highest EQE values for a two-terminal device without requiring any gate voltage, which can be beneficial for many imaging applications. This extremely large EQE suggests that our STMDS device has great potential for an extremely sensitive UV photodetector.

To understand the origin of this extraordinary EQE, we measured photoresponsivity behavior as a function of bias voltage and laser power. Fig.4b shows EQE as a function of bias voltage at 77 K while the device was illuminated by a 405 nm laser. The lower EQE value in Fig.4b is due to the high laser power used in this measurement. We observed a very linear behavior of EQE as a function of the bias voltage, which suggests that the gain mechanism is related to the photogating of the sample. The vanishing of photocurrent or EQE at $V_b = 0$ indicates that photovoltaic effect, which may occur for a metal/TMD interface,[63] does not contribute to any photocurrent in our devices. We attribute the absence of the photovoltaic effect to the absence of the Schottky barrier at the semimetal/$MoS_2$ interface. To determine the photoresponse mechanism further, we have done a laser power-dependent photocurrent study as shown in Fig.4c.

Following the earlier literature,[53, 59, 64] we have analyzed our laser power-dependent data using the Hornbeck-Haynes model.[65, 66] It has been demonstrated that the structural defects and disorder cause the band tail states or shallow trap states near the valence band and conduction band.[67-69] In addition to the shallow trap states, there also exists deep recombination centers, also known as midgap states, which cause nonradiative (Shockley−Read−Hall-type) recombination.[53, 59] The physical mechanism is shown schematically in the inset of Fig.4c. For the $n$-doped 1L-MoS$_2$, only the hole traps near the valence band are relevant. The trapping and de-trapping of the hole states occur with rate $1/\tau_t$ and $1/\tau_d$, respectively. If the trap states cause the electrostatic gating, it will shift the Fermi energy and increase the electrical conduction. Using this model, the photogated current is given by (see Supporting Information for detail calculations),

$$I_{\mathrm{PC}} = A \frac{1}{1 + \frac{B}{P_D}}$$

The two parameters $A$ and $B$ are given by

$$A = \frac{D_t C_g C_q}{e(C_g + C_q)} \frac{dV_b}{dV_g}; \qquad B = \frac{D_t hc}{\eta \lambda \tau_r} \left( \frac{\tau_t}{\tau_d} \right)$$

where $D_t$ is the density of the localized traps, $C_g$ is the geometrical capacitance, $C_q$ is the quantum capacitance, $V_g$ is the gate voltage, $e$ is the electron charge, $h$ is the Planck constant, $\lambda$ is the excitation laser wavelength, and $\eta$ is the absorption coefficient of 1-MoS$_2$. The laser power-dependent photocurrent is fitted with this model using $A$ and $B$ as the fitting parameter as shown in Fig.4c. We obtained excellent fitting of the experimental data confirming that photogating is the main mechanism in our devices. Since we have used an undoped substrate, we couldn't determine the $\frac{dV_b}{dV_g}$ and the density of traps using the fitting parameters.

Fig.4d presents EQE as a function of laser power measured at 77 K. Note that EQE decreases logarithmically as a function of laser power, which is also a signature of the photogating effect.[61]

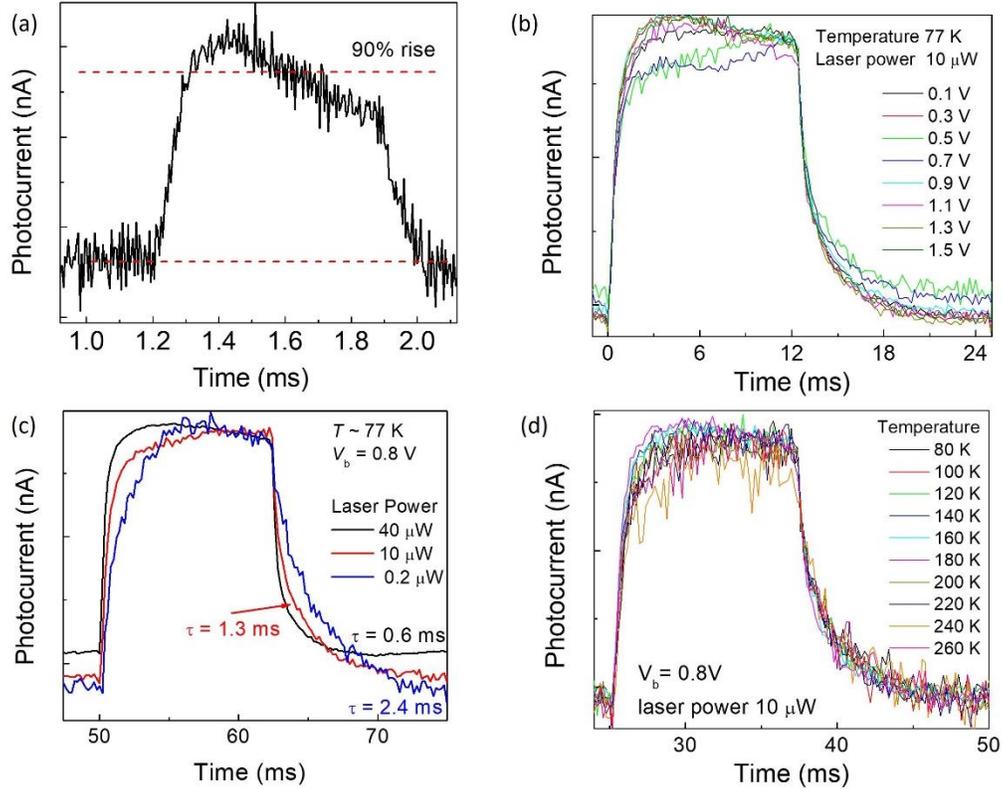

Figure 5: Time response of the device measured at 77K. (a) Time response of the device for a single laser pulse. We used a 405 nm laser modulated by a mechanical chopper ($f \sim 700$ Hz). We observed that a 90% rise time is 0.1 ms. (b) Time response of the device for different $V_b$ measured for a 10 μW laser power. No correlations between the decay time after the cessation of the laser and the bias voltage has been observed. (c) Time response as a function of different laser power. The measured decay time increases as we increase the laser power. The determined decay time is marked next to the lines. (d) Time response as a function of temperature. No correlations between the decay time after the cessation of the laser and the temperature has been observed.

Finally, we studied the time response of our devices at 77 K as shown in Fig.5. The devices were illuminated by a chopped laser ($f = 700$ Hz) and the signal was measured by an oscilloscope. The bias voltage of the devices was 2V. The time response of the photocurrent for a single pulse is shown in Fig.5a. We observed that the 90% rise time of the photocurrent is 0.1 ms, i.e., the frequency response of the current devices is ~ 10 kHz as shown in Fig.5a. By measuring photocurrent for a long time after terminating the laser illumination, we didn't observe the presence of persistent photocurrent in our STMDS devices (see Supporting Information).

To understand the decay time of the photocurrent in the dark, we also studied time response as a function of $V_b$, laser power, and temperature $T$, as shown in Fig.5b-d. We observed that the decay time in the dark does not depend on temperature and the bias voltage as shown in Fig.5b and Fig.5d, respectively. On the other hand, the decay of the photocurrent after cessation of laser excitation depends strongly on the laser power impinging on the device as shown in Fig.5c. The decay time was measured by fitting an exponential decay function ($I = I_0 e^{-t/\tau}$, where $t$ is the time and $\tau$ is the decay constant). We observed that the decay constant decreases monotonically as we increase the laser power. This is also a signature of the photogating effect. At a lower intensity, the trap states remain unsaturated and dominate the photocurrent decay after the cessation of laser excitation. With increasing laser power, the trap states get saturated and don't dominate the photocurrent decay resulting in a much lower (i.e., faster) decay time. Similar property of decay time as a function of control gate voltage was reported for a 1L MoS$_2$ phototransistor.[58] Hence, all our results consistently suggest that the high EQE observed in our devices are originating from the photogating effect in 1L MoS$_2$ or at the interface between 1L MoS$_2$ and the oxide layer.

The semimetal contacts to 1L-MoS$_2$ is playing a crucial role in obtaining the high photoresponsivity in our devices. The vanishing of Schottky barriers and low contact resistances due to semimetal/MoS$_2$ interface allow high injection of current when photogenerated carriers lowers the conductivity of the sample.

An interesting photodetector device structure would be to investigate devices with a Bi/1L-MoS$_2$/Au devices so that one contact is Ohmic and the other contact is Schottky contact. We plan to extend our investigation by studying those devices and present the result in a future publication.

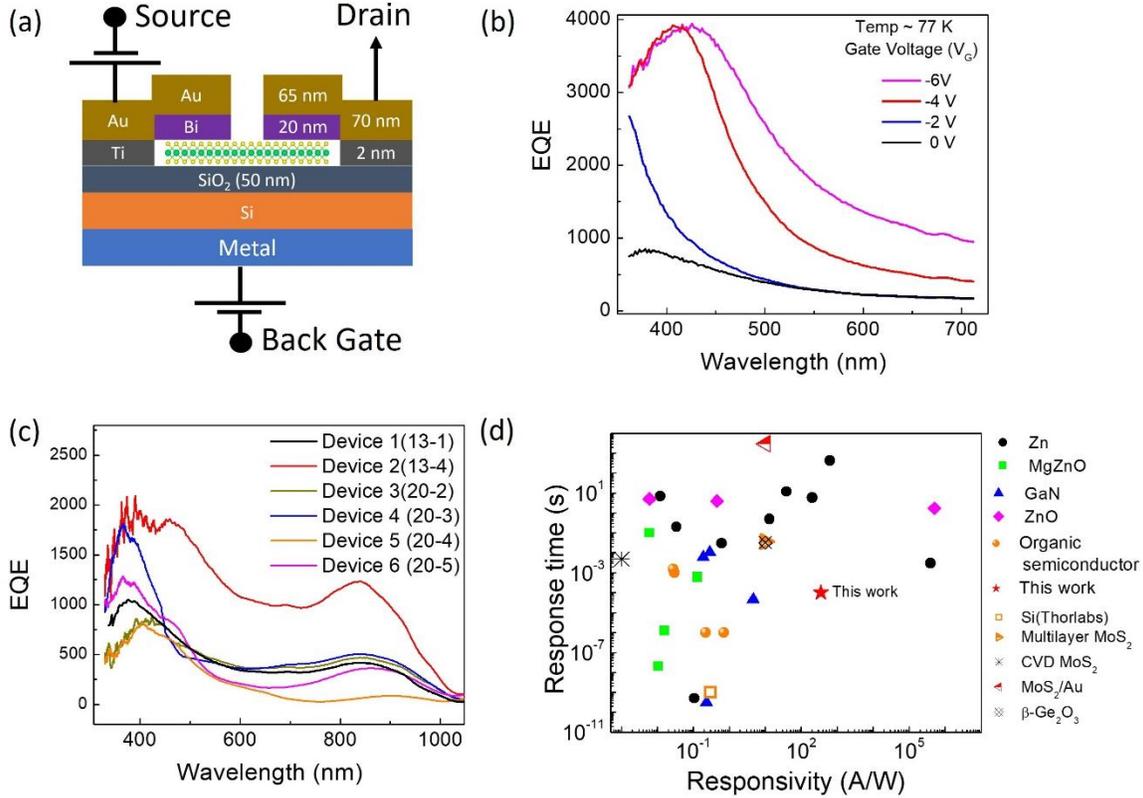

Figure 6: (a) Circuit configuration of an STMDS device as a phototransistor. This circuit configuration was used to study the electrostatic gating effect on photocurrent. (b) EQE of the device for back-gate voltages 0V (black), -2V (blue), -4V(red), and -6V (magenta). (c) External quantum efficiency of six devices measured at 77 K with a bias voltage $V_b = 2$ V. (d) Responsivity values vs response time plot for different type of solid-state UV photodetectors reported in the literature. See main text for the references.

To tune the EQE by using electrostatic gating, we measured photocurrent EQE of our STMDS devices by applying a back-gate voltage. We used Silver Conductive Paint or Silver Colloidal Suspension on the back side of the Si substrate to prepare a metallic gate. The circuit configuration of an STMDS device with a back-gate is shown in Fig. 6a, which is effectively working as a phototransistor. We found that EQE can be enhanced significantly by electrostatically doping the samples as shown in Fig.6b. The EQE was enhanced by 4 times ($EQE_{Max} \sim$ 4000) when the back-gate voltage is -6 V. This also confirms that the high EQE observed in our devices is due to photogating effect.

We present the external quantum efficiency plots for six different devices as shown in Fig. 6(a) to demonstrate the device performance variation. All the measurements were conducted under the same optical and electrical settings at 77 K. We found that the maximum responsivity at $\lambda \sim 400$ nm varies by ~200% from the lowest values as shown in Fig. 6(a).

Now we will discuss the figure of merits of our UV photodetector compared to the solid-state UV photodetectors reported in the literature. The responsivity and the response time of different solid-state UV photodetectors are shown in Fig.6b, which are obtained from the review article by Alaie et al.[1] The responsivity and the response-time data for multilayer $MoS_2/Au$, CVD $MoS_2/Au$ and exfoliated 1L, exfoliate $MoS_2/Au$, and $\beta Ge_2O_3$ devices were reported by Zhang et al.,[70] Yore et al,[46] Furchi et la.[53], and Arora et al.,[71] respectively. Fig.6b clearly shows that our semimetal-TMD based UV photodetector demonstrates superior performance. Hence, semimetal-TMD-semimetal UV photodetector possesses a true potential for next-generation solid-state UV photodetector.

Because of outstanding structural and mechanical properties,[72] monolayer TMDs can be easily fabricated on Si-based photonic structures for many applications ranging from covert communications, and biological analysis to fire monitoring and UV astronomy. Fast response makes STMDS devices attractive for optical to electrical interconnects, which may find applications in communication devices.[73, 74] High EQE values of STMDS make them attractive for low-UV light-level detections, such as single UV photon applications or photon-starved UV astronomy.

## 3. Conclusion

In conclusion, we demonstrated a large array fabrication of fast and ultrasensitive photodetectors based on CVD-grown 1L $MoS_2$ electrically connected by a semimetal that forms an ohmic contact at the interfaces and demonstrates efficient photodetection. We determined several important figures of merits for our devices: responsivity, external quantum efficiency, time response, and scalability. Our results yield a fundamental understanding of semimetal-TMD-semimetal devices and will provide important information to develop next-generation TMD-based nanophotonic devices.

**Acknowledgment:**

We thank Prof. Krishna Saraswat for stimulating discussion. A.K.M.N and H.S. acknowledge the support from the Department of Defense Award (ID: W911NF-19-1-0007) and NSF Award (ECCS- 2151971). A.K.M.N. also acknowledges the faculty start-up grant, Kenneth S. Fong Translational Award, and small grant provided by the College of Science and Engineering at San Francisco State University. A.K. and E.P. acknowledge support from Stanford SystemX alliance and Samsung Semiconductor. Part of this work was performed at the Stanford Nanofabrication Facility (SNF), supported by the National Science Foundation under award ECCS-2026822.

**Supporting Information**

Supporting Information is available from the Wiley Online Library or from the author.

Supporting Information

Semimetal-Monolayer Transition Metal Dichalcogenides Photodetectors for

Wafer-Scale Broadband Photonics

Hon-Loen Sinn,[1] Aravindh Kumar,[2] Eric Pop,[2,3,4] and Akm Newaz[1]

[1]Department of Physics and Astronomy, San Francisco State University, San Francisco, California 94132, USA

[2]Department of Electrical Engineering, Stanford University, Stanford, California 94305, USA

[3]Department of Materials Science and Engineering, Stanford University, Stanford, California 94305, USA

[4]Precourt Institute for Energy, Stanford University, Stanford, California 94305, USA

## S1. Density of states and the band Structure of metal/semiconductor and semimetal/semiconductor

We propose a mechanism, similar to the mechanism proposed by Shen *et al.*,[1] that the alignment of the Fermi level, near-zero DOS of a semimetal, and the bottom of the conduction-band minima is the principle reason for the formation of the ohmic contact at the interface. We presented the density of states and the corresponding band structure schematically in Fig.S1, which is recreated from the work by Shen et al.[1] For a semimetal/semiconductor interface, the Fermi level is located at the near-zero density of states (DOS) and aligned with the bottom of the conduction band. Hence, the MIGS from the conduction-band tail is heavily suppressed and the contact become Schottky barrier free.

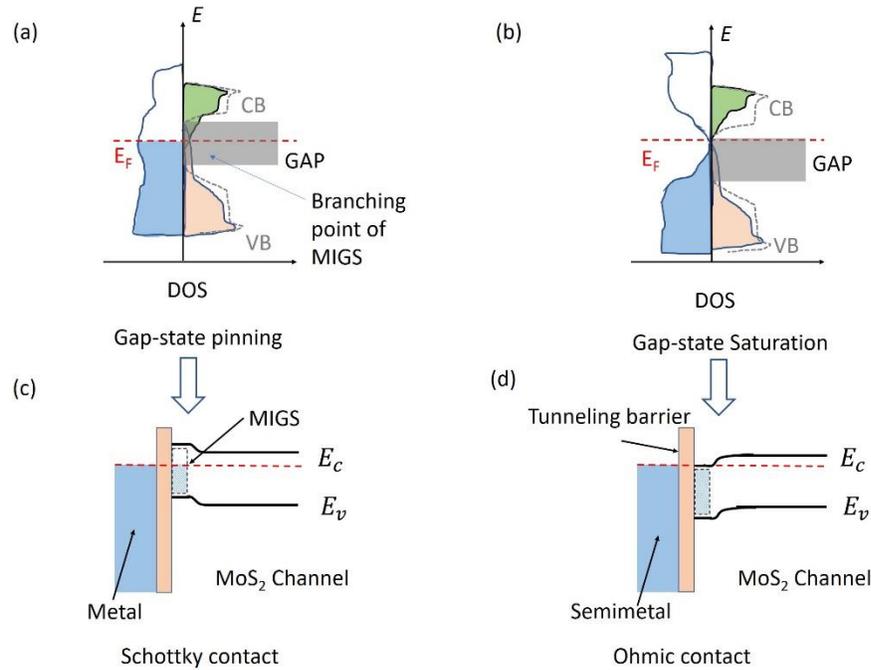

Figure S1: (a)-(df) Schematically drawn band structure for metal/1L-MoS2 and semimetal/1L-MoS2 contacts. This figure is recreated from the work by Shen *et al.*[1] (a) The figure shows the density of states (DOS) of normal metal and semiconductor contact. The light green (light red) shaded are represents the contributions of the conduction band (CB) (valence band (VB)) to the MIGS state. Since the Fermi level ($E_F$) is located at the branching point of the MIGS, the contact become a Schottky type contact. The light blue shaded area in the metal side present the electron occupied states. (b) This figure shows the DOS of semimetal and semiconductor contact. Since, the Fermi level is at the near-zero DOS and is aligned with the conduction band minima of the semiconductor, the MIGS are suppressed and the contact become ohmic. (c) The band structure of a normal metal/semiconductor contact. The Fermi level pinning at the MIGS causes a Schottky barrier at the contact. (d) The band structure of a semimetal/semiconductor contact. The gap state saturation at the contact causes an ohmic contact.

## S2. Carrier dynamics

We calculated the photo-response using a modified Hornback-Haynes model,[2-6] in which the valence band has a tail by a distribution of states with density $D_t$ and with energy $E_{v,t}$ above the valence band with Energy $E_v$. The photogenerated holes get trapped and de-trapped in those trap states near the valence band. Since our device is ON state due to the n-doping and as the Fermi energy is near the edge of the conduction band, we neglected the electron trapping. The recombination occurs via midgap states with an empirical (constant) rate $1/\tau_r$. We assumed that the recombination rate for both electrons and holes is the same. The holes are trapping into the states and de-trapping out of the states with a rate $1/\tau_t$ and $1/\tau_d$, respectively. We presented the physical mechanism schematically in Fig.S2.

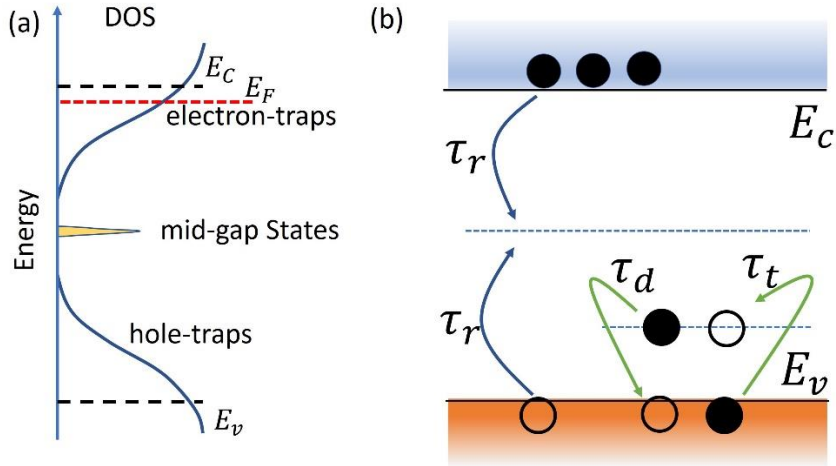

Figure S2: (a) Schematic illustration of the density-of-states (DOS) in 1L-MoS₂. The conduction band and valence band are at energy $E_C$ and $E_v$, respectively. Band tail states exist underneath (above) the conduction (valence) band edge that act as electron (hole) charge traps. The recombination to occur via midgap states with an empirical (constant) rate $1/\tau_r$. We have assumed that the recombination rate for both electron and holes are the same. (b) Simplified energy band diagram that shows the main features of the charge trapping and detrapping model. The valence band trail is approximated by a discrete distribution of hole traps with density $D_t$ (occupation of traps $p_t$). The holes are trapping into the states and de-trapping out of the sates with a rate $1/\tau_t$ and $1/\tau_d$, respectively.

Optical illumination creates carrier density in 1L-MoS₂. Because of charge neutrality, the change of carrier density due to optical illumination is given by,

$$\Delta n = \Delta p + p_t \qquad (1)$$

where $\Delta n$ ($\Delta$p) is the free electron (hole) concentration and $p_t$ is the trapped hole carrier concentrations. The change in conductivity due to the optical illumination is given by,

$$\Delta \sigma = q(\mu_n + \mu_p)\Delta p + q\mu_n p_t \qquad (2)$$

where $\mu_n(\mu_p)$ are the electron(hole) mobility and $q$ is the electron charge. Hence the change in the conduction due to the presence of the traps is $q\mu_n p_t$.

Now, the equations for the carrier dynamics are given by,

$$\frac{d\Delta p}{dt} = \varphi - \frac{\Delta p}{\tau_r} + \frac{p_t}{\tau_d} - \frac{\Delta p}{\tau_t}\left(1 - \frac{p_t}{D_t}\right)\text{---------------}(3)$$

$$\frac{dp_t}{dt} = -\frac{p_t}{\tau_d} - \frac{\Delta p}{\tau_t}\left(1 - \frac{p_t}{D_t}\right)\text{----------------------}(4)$$

where $\varphi$ is the absorbed photon flux and is given by $\eta P_D \lambda/hc$. Here $\eta$ is the photon absorption coefficient, $P_D$ is the optical power density, $\lambda$ is the wavelength, $h$ is the Planck constant, and $c$ is the speed of light.

The steady-state condition allows us to set $\frac{d\Delta p}{dt} = \frac{dp_t}{dt} = 0$. By applying this steady condition, we obtain the following two equations.

$$\Delta p = \varphi\tau_r\text{---------------------}(5)$$

$$p_t = \frac{\varphi D_t \tau_t}{\varphi\tau_r + D_t\left(\frac{\tau_t}{\tau_d}\right)}\text{----------------}(6)$$

Since there is $p_t$ holes are trapped, it creates an electrostatic voltage or gate voltage ($\Delta V_G$), which we calculated using a simple capacitor model.

$$\Delta V_G = \frac{qp_t}{C}\text{------------------}(7)$$

where $C$ is the effective capacitance and is given by,

$$\frac{1}{C} = \frac{1}{C_g} + \frac{1}{C_q}$$

where $C_g$ is the geometrical capacitance and $C_q$ the quantum capacitance defined as $C_q = e2g_{2D}$. Here $g_{2D}$ is the density of states of a 2D electron gas system and $e$ is the electron charge.

We can calculate the photocurrent as

$$I_{PC} = \Delta V_G \frac{dI_{ds}}{dV_G}\text{-------------}(8)$$

where $\frac{dI_{ds}}{dV_G}$ is the transverse conductance. By using Eq.8 and Eq.6, we can get

$$I_{PC} = \frac{qD_t}{C}\frac{dI_{ds}}{dV_g}\frac{1}{1 + \frac{D_t}{\varphi\tau_r}\left(\frac{\tau_t}{\tau_d}\right)} = A\frac{1}{1 + \frac{B}{P_D}}\text{----------}(9)$$

where two new parameters $A$ and $B$ are given by,

$$A = \frac{qD_t}{C}\frac{dI_{ds}}{dV_g}, \qquad B = \frac{D_t hc}{\eta\lambda\tau_r}\left(\frac{\tau_t}{\tau_d}\right) \qquad\qquad \text{----------}(10)$$

S4

## S3: Photoresponsivity of an Au/Ag/1L-MoS₂/Ag/Au device

We studied CVD grown devices that are connected with Ag/Au electrical contacts. The devices were grown directly on a glass substrate. Figure S3 shows the optical image of one such device along with the photoresponsivity measured at 77 K. We observed that the photoresponsivity in Ti/Au connected device is six orders of magnitude lower than Bi- contacted device.

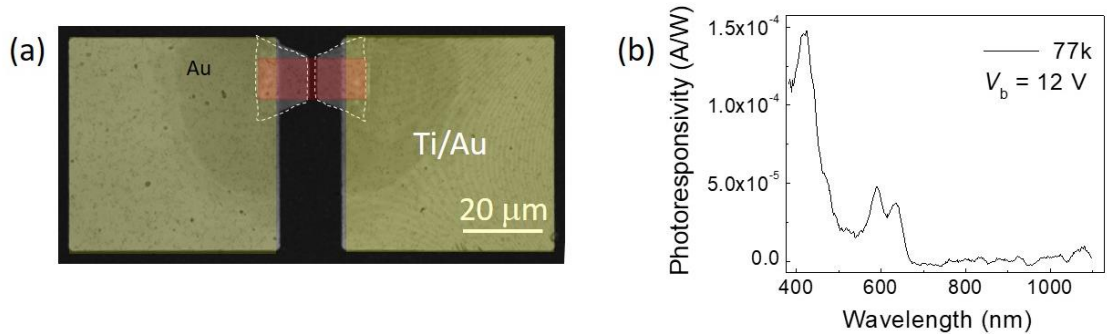

Figure S3: (a) The optical image (False-colored) of the sample. The 1L-MoS₂ etched ribbon is marked by a red rectangle. The Ag/Au (5/25 nm) electrical connection to the sample is marked by dashed trapezoid. (b) The photoresponsivity of sample measured at 77 K. The applied bias voltage was 12 V. We observed three peaks associated with the A-, B-, and C- excitons. We observed zero photocurrent for wavelengths above 700nm.

## S4: Scanning photocurrent microscopy image

Scanning photocurrent image of a sample is shown Fig.S4. Since our sample resides inside a cryostat and we use a long working distance microscope objective (working distance ~ 17 mm), our laser beam diameter (~ 2 μm) is larger than the diffraction limited size. The scanning photocurrent image was measured at 77 K. We didn't observe any photocurrent outside MoS₂.

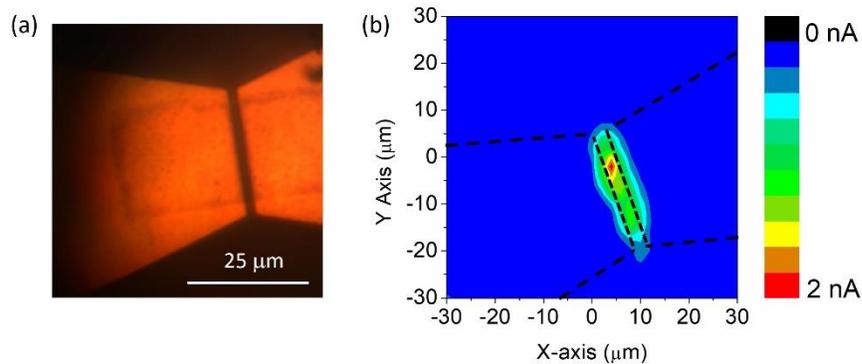

Figure S4: (a) The optical image of the sample used for the scanning photocurrent measurements. (b) The figure presents the scanning photocurrent measured at 77 K. The dashed black lines present the outline of the metal pads.

## S5: Persistent photocurrent study

TMD based devices demonstrated persistent photoconductivity (PPC), which is sustained conductivity after illumination is blocked or removed.[7] The PPC in MoS2 has been attributed to the charge traps due to the inhomogeneities in the substrate. To investigate PPC, we have studied for 5 minutes after terminating the laser illumination as shown in Fig.S5. No persistent photocurrent was observed. Interestingly, we found that the dark current has some transient behavior whose origin is not currently known.

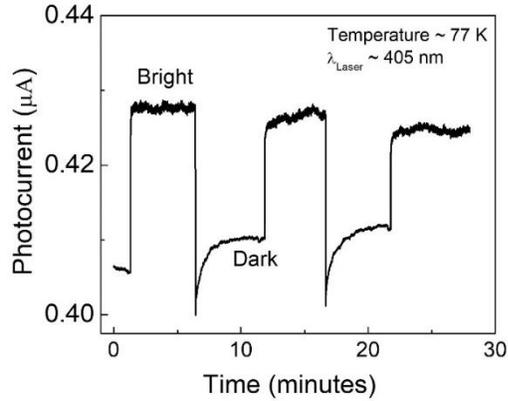

Figure S5: Photocurrent as we illuminate the sample and terminate the illumination. The measurement was conducted at 77 K. The photocurrent was measured by a digital multimeter (Keithley 2000).

**S6. Time response study of a Bi-MoS₂-Bi device using a red laser**

We investigated the time response as a function of $V_b$, laser power, and temperature $T$, for a red laser (wavelength, $\lambda$ =650 nm) as shown in Fig.6a-d.

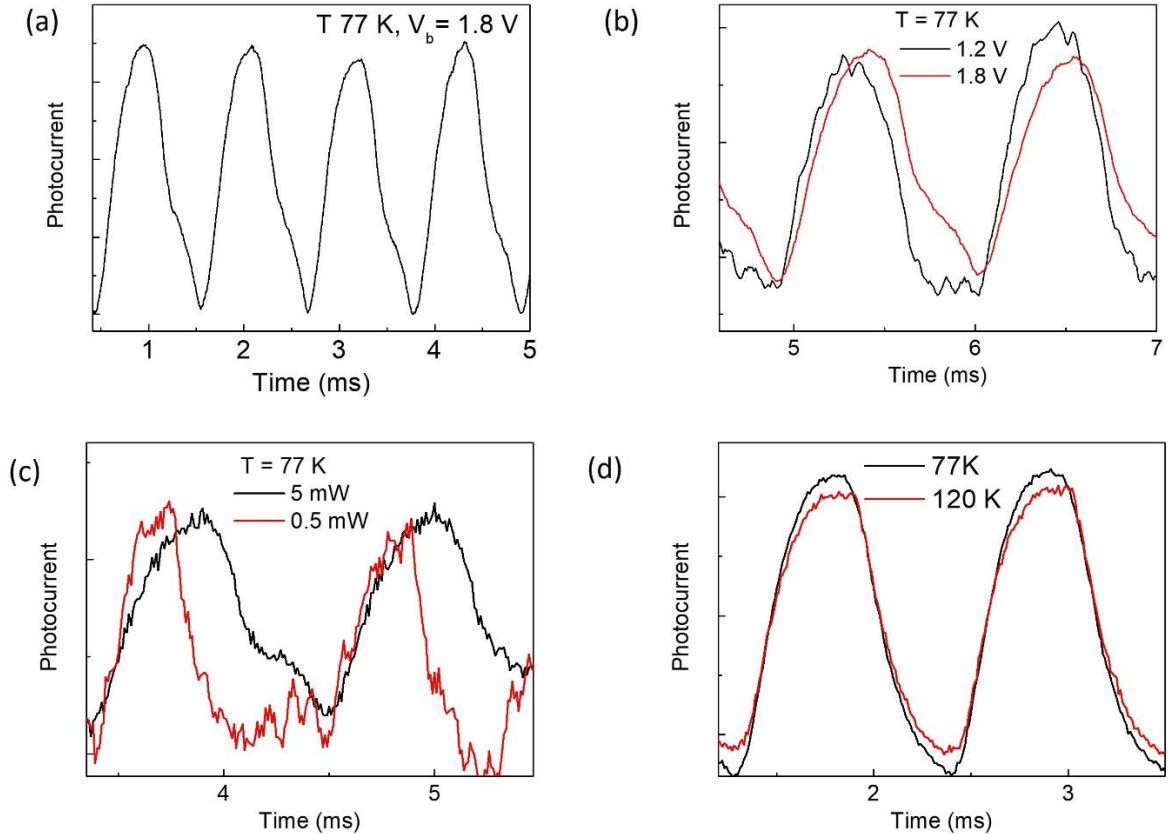

Figure S6: Time response of the device measured at 77K. (a) Time response of the device for a single laser pulse. We used a 650 nm laser modulated by a mechanical chopper ($f$ ~ 700 Hz). We observed that a 90% rise time is 0.4 ms. (b) Time response of the device for different $V_b$ measured for a 50 μW laser power. No correlations between the decay time after the cessation of the laser and the bias voltage has been observed. (c) Time response as a function of different laser power. The measured decay time decreases as we increase the laser power. (d) Time response as a function of temperature. No correlations between the decay time after the cessation of the laser and the temperature has been observed.